\documentclass[hyper,letterpaper,12pt]{article}
\usepackage{a4wide}
\usepackage{amsmath}
\usepackage[
      colorlinks=true,
      linkcolor=blue,
      urlcolor=blue,
      filecolor=black,
      citecolor=blue,
            ]{hyperref}
\usepackage{color}
\usepackage{graphicx}
\usepackage{amsfonts}
\usepackage{amssymb}
\usepackage{epstopdf}

\def\beq{\begin{equation}}
\def\eeq{\end{equation}}

\begin{document}
\title{\bf \Large A Holographic P-wave Superconductor Model}

\author{\large
~Rong-Gen Cai$^1$\footnote{E-mail: cairg@itp.ac.cn}~,
~~Li Li$^1$\footnote{E-mail: liliphy@itp.ac.cn}~,
~~Li-Fang Li$^2$\footnote{E-mail: lilf@itp.ac.cn}\\
\\
\small $^1$State Key Laboratory of Theoretical Physics,\\
\small Institute of Theoretical Physics, Chinese Academy of Sciences,\\
\small Beijing 100190,  China.\\
\small $^2$State Key Laboratory of Space Weather, \\
\small Center for Space Science and Applied Research, Chinese Academy of Sciences,\\
\small Beijing 100190, China.}
\date{\today}
\maketitle

\begin{abstract}
\normalsize We study a holographic p-wave superconductor model in a four dimensional Einstein-Maxwell-complex vector field theory with a negative cosmological constant. The complex vector field is charged under the Maxwell field. We solve the full coupled equations of motion of the system and find  black hole solutions with the vector hair. The vector hairy black hole solutions are dual to a thermal state with the U(1) symmetry as well as the spatial rotational symmetry broken spontaneously. Depending on two parameters, the mass and charge of the vector field, we find a rich phase structure: zeroth order, first order and second order phase transitions can happen in this model. We also find ``retrograde condensation" in which the hairy black hole solution exists only for the temperatures above a critical value with the free energy much larger than the one of the black hole without the vector hair. We construct the phase diagram for this system in terms of the temperature and charge of the vector field.
\end{abstract}

\tableofcontents

\section{ Introduction}

Due to the strong/weak duality characteristic of the Anti-de
Sitter/Conformal Field  Theory correspondence
(AdS/CFT)~\cite{Maldacena:1997re,Gubser:1998bc,Witten:1998qj}, it
provides us with a powerful approach to study the properties of
strong coupled systems by a weak coupled AdS gravity. The high
temperature superconductivity is a potential area where the AdS/CFT
correspondence is applicable. According to the symmetry of the
spatial part of wave function of the Cooper pair, superconductors
can be classified as the s-wave, p-wave, d-wave, f-wave
superconductor, etc. From a phenomenological perspective, the onset
of superconductivity is characterized by the condensation of a
composite charged operator spontaneously breaking U(1) symmetry at
some temperature. The holographic s-wave superconductor model was
first realized in refs.~\cite{Hartnoll:2008vx,Hartnoll:2008kx}.
According to the AdS/CFT correspondence, in the gravity side, a
Maxwell field and a charged scalar field are introduced to describe
the U(1) symmetry and the scalar operator in the dual field theory
side. This holographic model undergoes a phase transition from black
hole with no hair (normal phase/conductor phase) to the case with
scalar hair at low temperatures (superconducting phase). Holographic
d-wave model was constructed by introducing a charged massive spin
two field propagating in the
bulk~\cite{Chen:2010mk,Benini:2010pr,Kim:2013oba}. To realize a
holographic p-wave superconductor model, one needs to introduce a
charged vector field in the bulk as a vector order parameter.
Ref.~\cite{Gubser:2008wv} presented a holographic p-wave model by
introducing a SU(2) Yang-Mills field into the bulk, where a gauge
boson generated by one SU(2) generator is dual to the vector order
parameter. Other generalized studies based on this model can be
found for example in
refs.~\cite{Roberts,Zeng:2010fs,Cai:2010zm,Zayas:2011dw,Momeni:2012ab,Roychowdhury:2013aua}.
An alternative holographic realization of p-wave superconductivity
emerges from the condensation of a 2-form field in the
bulk~\cite{Aprile:2010ge}.

In a recent paper~\cite{Cai:2013pda}, we have studied a holographic model by introducing a complex vector field $\rho_\mu$ charged under a Maxwell gauge field $A_\mu$ in the bulk, which is dual to a strongly coupled system involving a charged vector operator with a global U(1) symmetry. In this model there exists a non-minimal coupling between the vector field and the gauge field characterizing the magnetic moment of the vector field, which plays a crucial role in the condensate of the vector field induced by an applied magnetic field. We have studied this model in the probe limit at finite density. Such a setup meets the minimum requirement to construct a holographic p-wave superconductor model. Indeed, we have found a critical temperature at which the system undergoes a second order phase transition. The critical exponent of this transition is one half which coincides with the case in the Landau-Ginzburg theory. In the condensed phase, a vector operator acquires a vacuum expectation value breaking the U(1) symmetry as well as rotational symmetry spontaneously. Our calculation indicates that this condensed phase exhibits an infinite DC conductivity and a gap in the optical conductivity, which is very reminiscent of some characteristics known from ordinary superconductivity. In this sense, our model can be regarded as a holographic p-wave model.

The probe approximation neglecting the back reaction of the matter fields is only justified in the limit of large $q$ with $q\rho_\mu$ and $qA_\mu$ fixed. It has been shown that new phases can emerge (see refs.~\cite{Cai:2013wma,Liu:2013yaa,Nitti:2013xaa} for example) and the order of the phase transition can also be  changed~\cite{Horowitz:2010jq,Ammon:2009xh,Peng:2011gh,Cai:2012es,Cai:2013oma} once the back reaction of the matter fields on the geometry is taken into account. To study the complete phase diagram of our holographic system, we need to go beyond the probe approximation and to include the back reaction. While the previous paper~\cite{Cai:2013pda} focused on the effects of the non-minimal coupling term and applied magnetic field on the condensate of the vector operator, in this paper we aim at studying the effect of the back reaction of the matter fields on the background geometry. We will turn off the non-minimal coupling between the vector field $\rho_\mu$ and the gauge field $A_\mu$ since we do not discuss magnetic effect in this paper. So the model is left with two independent parameters, i.e., the mass $m$ of the vector field giving the dimension of the dual vector operator and its charge $q$ controlling the strength of the back reaction on the background geometry. We manage to construct asymptotically AdS charged black hole solutions with nontrivial vector hair. It turns out that depending on $m^2$ and $q$, our model exhibits a rich phase structure.

The thermodynamic behavior of the model has a dramatic change from large $m^2$ to small $m^2$. In the case with large $m^2$, if one lowers the temperature, the normal phase will become unstable to developing vector hair below a critical temperature $T_c$. The transition from the normal phase to the condensed phase is second order for larger $q$, i.e., weak strength of the back reaction. However, as we decrease $q$ to a critical one, the phase transition becomes first order. On the other hand, for the case with small $m^2$, no matter the value of $q$, there exists a temperature below which the condensed phase never exists. When the back reaction is weak, hairy solutions dominate the phase diagram below a critical temperature $T_2$ through a second order transition, then the condensed phase terminates at a lower temperature $T_0$ at which its free energy jumps to the one in the normal phase, indicating a zeroth order transition. As we strengthen the back reaction, we first encounter for a first order transition at temperature $T_1$ and then a zeroth order transition at $T_0$. For the sufficiently strong back reaction
case, the condensed phase only occurs at a high temperature $T>T_n$ rather than at a low temperature. Furthermore, the hairy phase has higher free energy than the normal phase. The four critical transition temperatures $T_c$, $T_2$, $T_1$ and $T_0$ decrease as one increases the strength of the back reaction. To summarize possible phases associated with different ranges of model parameters, we construct the phase diagram in terms of charge $q$ and temperature $T$ for a given mass. We find that the critical temperature increases with the charge and decreases with the mass of $\rho_\mu$.

This paper is organized as follows. In the next section, we introduce the holographic model and deduce the equations of motion of the model. In section~\ref{sect:motion}, we give our ansatz for the hairy black hole solution corresponding to the condensed phase and specify the boundary conditions to be satisfied. Section~\ref{sect:free} is devoted to calculating the free energy and dual stress-energy tensor. We present numerical results in section~\ref{sect:superconducor}. For each given $m^2$, we scan a wide range of $q$ to find all possible types of phase transitions and construct the phase diagram. The conclusion and some discussions are included in section~\ref{sect:conclusion}.


\section{The holographic model}
\label{sect:model}

Let us introduce a complex vector field $\rho_\mu$, with mass $m$ and charge $q$, into the $(3+1)$ dimensional Einstein-Maxwell theory with a negative cosmological constant. The complete action reads
\begin{equation}\label{action}
\begin{split}
S=\frac{1}{2\kappa^2}\int d^4 x
\sqrt{-g}(\mathcal{R}+\frac{6}{L^2}+\mathcal{L}_m),\\
\mathcal{L}_m=-\frac{1}{4}F_{\mu\nu} F^{\mu \nu}-\frac{1}{2}\rho_{\mu\nu}^\dagger\rho^{\mu\nu}-m^2\rho_\mu^\dagger\rho^\mu+iq\gamma \rho_\mu\rho_\nu^\dagger F^{\mu\nu},
\end{split}
\end{equation}
with $L$ the AdS radius set to be unity and $\kappa^2\equiv 8\pi G $ related to the gravitational constant in the bulk. The Maxwell field strength reads $F_{\mu\nu}=\nabla_\mu A_\nu-\nabla_\nu A_\mu$. $\rho_{\mu\nu}$ in~\eqref{action} is defined by $\rho_{\mu\nu}=D_\mu\rho_\nu-D_\nu\rho_\mu$ with the covariant derivative $D_\mu=\nabla_\mu-iq A_\mu$. The last non-minimal coupling term characterizes the magnetic moment of the vector field $\rho_\mu$, which plays an important role in the case with an applied magnetic field~\cite{Cai:2013pda}. In the present study, since we only consider the case without external magnetic field, this term will not play any role.

Varying the action~\eqref{action}, we obtain the equations of motion for matter fields
\begin{eqnarray}
\label{gauge}
&& \nabla^\nu F_{\nu\mu}=iq(\rho^\nu\rho_{\nu\mu}^\dagger-{\rho^\nu}^\dagger\rho_{\nu\mu})+iq\gamma\nabla^\nu(\rho_\nu\rho_\mu^\dagger-\rho_\nu^\dagger\rho_\mu),
\\
\label{vector}
&& D^\nu\rho_{\nu\mu}-m^2\rho_\mu+iq\gamma\rho^\nu F_{\nu\mu}=0,
\end{eqnarray}
and the equations of gravitational field
\begin{equation}\label{tensor}
\begin{split}
\mathcal{R}_{\mu\nu}-\frac{1}{2}\mathcal{R}g_{\mu\nu}&-\frac{3}{L^2}g_{\mu\nu}=\frac{1}{2}F_{\mu\lambda}{F_\nu}^\lambda+\frac{1}{2}\mathcal{L}_m g_{\mu\nu}\\
&+\frac{1}{2}\{[\rho_{\mu\lambda}^\dagger{\rho_\nu}^\lambda+m^2{\rho_\mu}^\dagger\rho_\nu-iq\gamma(\rho_\mu{\rho_\lambda}^\dagger-{\rho_\mu}^\dagger\rho_\lambda){F_\nu}^\lambda]+\mu\leftrightarrow\nu\}.
\end{split}
\end{equation}

In the AdS/CFT correspondence, a hairy black hole with appropriate boundary conditions can be explained as a condensed phase of the dual field theory, while a black hole without hair is dual to an uncondensed phase (normal phase). In our case, since $\rho_\mu$ is charged under the U(1) gauge field, its dual operator will carry the same charge under this gauge symmetry and a vacuum expectation value of this operator will then trigger the U(1) symmetry breaking spontaneously. More precisely, we hope that this system would admit hairy black hole solutions at low temperatures, but no hair at high temperatures. Thus, the condensate of the dual vector operator will break the U(1) symmetry as well as the spatial rotational symmetry since the
condensate will pick out one direction as special. Therefore, viewing this vector field as an order parameter, the holographic model can be used to mimic a p-wave superconductor (superfluid) phase transition. This turns out to be true in the probe limit~\cite{Cai:2013pda}: when one lowers the temperature to a certain value, the normal background becomes unstable and a nontrivial vector hair $\rho_x$ appears. In this paper, we continue to study this model by considering the back reaction of matter fields on the background geometry.

\section{Equations of motion and boundary conditions}
\label{sect:motion}
To construct homogeneous charged black hole solutions with vector hair, we adopt the following ansatz
\begin{equation}\label{ansatz}
\begin{split}
ds^2=-f(r)e^{-\chi(r)}dt^2+\frac{dr^2}{f(r)}+r^2h(r)dx^2+r^2dy^2,\\
\rho_\nu dx^\nu=\rho_x(r)dx,\quad A_\nu dx^\nu=\phi(r)dt.
\end{split}
\end{equation}
We will denote the position of the horizon as $r_h$ and the conformal boundary will be at $r\rightarrow\infty$. Our consideration is as follows. Since we would like to study a dual theory with finite chemical potential or charge density accompanied by a U(1) symmetry, we turn on $A_t$ in the bulk. We want to allow for states with a non-trivial current $\langle\hat{J_x}\rangle$, for which we further introduce $\rho_x$ in the bulk. Because a non-vanishing $\langle\hat{J_x}\rangle$ picks out $x$ direction as special, which obviously breaks the rotational symmetry in $x-y$ plane. Therefore we introduce a function $h(r)$ in the $xx$ component of the metric in order to describe the anisotropy.

The horizon $r_h$ is determined by $f(r_h)=0$. The temperature $T$ of the black hole is given by
\begin{equation}\label{temp}
T=\frac{f'(r_h)e^{-\chi(r_h)/2}}{4\pi},
\end{equation}
and the thermal entropy $S$ is given by the Bekenstein-Hawking entropy of the black hole
\begin{equation}\label{entropy}
S=\frac{2\pi}{\kappa^2}A=\frac{2\pi V_2}{\kappa^2}r_h^2\sqrt{h(r_h)},
\end{equation}
where $A$ denotes the area of the horizon and $V_2=\int dxdy$.

One finds that the $r$ component of~\eqref{gauge} implies that the phase of $\rho_x$ must be constant. Without loss of generality, we can take $\rho_x$ to be real. Then, the independent equations of motion in terms of the above ansatz are deduced as follows
\begin{equation}\label{eoms}
\begin{split}
\phi''+(\frac{h'}{2h}+\frac{\chi'}{2}+\frac{2}{r})\phi'-\frac{2q^2\rho_x^2}{r^2fh}\phi=0,\\
\rho_x''+(\frac{f'}{f}-\frac{h'}{2h}-\frac{\chi'}{2})\rho_x'+\frac{e^{\chi}q^2\phi^2}{f^2}\rho_x-\frac{m^2}{f}\rho_x=0, \\
\chi'-\frac{2f'}{f}-\frac{h'}{h}+\frac{\rho_x'^2}{rh}-\frac{re^\chi\phi'^2}{2f}-\frac{e^\chi q^2\rho_x^2\phi^2}{rf^2h}+\frac{6r}{L^2f}-\frac{2}{r}=0,\\
h''+(\frac{f'}{f}-\frac{h'}{2h}-\frac{\chi'}{2}+\frac{2}{r})h'+\frac{2{\rho_x'}^2}{r^2}-\frac{2e^\chi q^2\rho_x^2\phi^2}{r^2f^2}+\frac{2m^2\rho_x^2}{r^2f}=0,\\
(\frac{2}{r}-\frac{h'}{2h})\frac{f'}{f}+(\frac{1}{r}+\frac{\chi'}{2})\frac{h'}{h}-\frac{\rho_x'^2}{r^2h}+\frac{e^\chi\phi'^2}{2f}+\frac{3e^\chi q^2\rho_x^2\phi^2}{r^2f^2h}-\frac{m^2\rho_x^2}{r^2fh}-\frac{6}{L^2f}+\frac{2}{r^2}=0,
\end{split}
\end{equation}
where the prime denotes the derivative with respect to $r$.

The full coupled equations of motion do not admit an analytical solution with non-trivial $\rho_x$. Therefore, we have to solve them numerically. We will use shooting method to solve
equations~\eqref{eoms}. In order to find the solutions for all the five functions $\mathcal{F}=\{\rho_x,\phi,f,h,\chi\}$ one must impose suitable boundary conditions at both conformal boundary $r\rightarrow\infty$ and the horizon $r=r_h$.

In order to match the asymptotical AdS boundary, the general falloff near the boundary $r\rightarrow\infty$ behaves as
\begin{equation} \label{boundary}
\begin{split}
\phi=\mu-\frac{\rho}{r}+\ldots,\quad \rho_x=\frac{{\rho_x}_-}{r^{{\Delta}_-}}+\frac{{\rho_x}_{+}}{r^{{\Delta}_+}}+\ldots,\\
f=r^2(1+\frac{f_3}{r^3})+\ldots,\quad h=1+\frac{h_3}{r^3}+\ldots,\quad \chi=0+\frac{\chi_3}{r^3}+\ldots,
\end{split}
\end{equation}
where the dots stand for the higher order terms in the expansion in
power of $1/r$ and ${\Delta}_\pm=\frac{1\pm\sqrt{1+4
m^2}}{2}$.~\footnote{The $m^2$ has a lower bound as $m^2=-1/4$ with
${\Delta}_+={\Delta}_-=1/2$. In that case, there is a logarithmic
term in the asymptotical expansion. We treat such a term as the
source set to be zero to avoid the instability induced by this
term~\cite{Horowitz:2008bn}. The treatment for $m^2=-1/4$ is very
subtle.  We will not discuss this case in this paper and instead we
are going to give a detailed study in future.} We impose
${\rho_x}_-=0$ since we want the condensate to arise spontaneously.
According to the AdS/CFT dictionary, up to a normalization, the
coefficients $\mu$, $\rho$, ${\rho_x}_{+}$ are regarded as chemical
potential, charge density and the $x$ component of the vacuum
expectation of the vector operator $\hat{J^\mu}$ in the dual field
theory, respectively.

We are interested in black hole configurations that have a regular event horizon located at $r_h$. Therefore,
 in addition to $f(r_h)=0$, one must require $\phi(r_h)=0$ in order for $g^{\mu\nu}A_\mu A_\nu$ being finite at the horizon.
 We require the regularity conditions at the horizon $r=r_h$, which means that all our functions have finite values and admit a series expansion in terms of $(r-r_h)$ as
\begin{equation}\label{series}
\mathcal{F}=\mathcal{F}(r_h)+\mathcal{F}'(r_h)(r-r_h)+\cdots.
\end{equation}
By plugging the expansion~\eqref{series} into~\eqref{eoms}, one can
find that  there are five independent parameters at the horizon
$\{r_h,\rho_x(r_h),\phi'(r_h),h(r_h),\chi(r_h)\}$. However, there
are three useful scaling symmetries in the equations of motion,
which read
\begin{equation} \label{scaling1}
e^\chi\rightarrow\lambda^2 e^\chi,\quad t\rightarrow\lambda t,\quad \phi\rightarrow\lambda^{-1}\phi,
\end{equation}
\begin{equation} \label{scaling2}
\rho_x\rightarrow\lambda \rho_x,\quad x\rightarrow\lambda^{-1} x,\quad h\rightarrow\lambda^2 h,
\end{equation}
and
\begin{equation} \label{scaling3}
r\rightarrow\lambda r,\quad (t,x,y)\rightarrow{\lambda^{-1}}(t,x,y),\quad(\phi,\rho_x)\rightarrow\lambda(\phi,\rho_x), \quad f\rightarrow\lambda^2f,
\end{equation}
where in each case $\lambda$ is a real positive number.

Taking advantage of above three scaling symmetries, we can first
set $\{r_h=1,\chi(r_h)=0, h(r_h)=1\}$ for performing numerics. After
solving the coupled differential equations, we should use the first
two symmetries again to satisfy the asymptotic conditions
$\chi(\infty)=0$ and $h(\infty)=1$. Thus we finally have two
independent parameters $\{\rho_x(r_h),\phi'(r_h)\}$ at hand. We
shall use $\phi'(r_h)$ as the shooting parameter to match the source
free condition, i.e., ${\rho_x}_-=0$. After solving the set of
equations, we can obtain the condensate $\langle \hat{J^x}\rangle$,
chemical potential $\mu$ and charge density $\rho$ by reading off
the corresponding coefficients in~\eqref{boundary}, respectively.

Under the third symmetry, the revelent quantities transform as
\begin{equation} \label{transform}
T\rightarrow\lambda T,\quad S\rightarrow S,\quad \mu \rightarrow\lambda\mu,\quad \rho\rightarrow\lambda^2\rho, \quad {\rho_x}_{+}\rightarrow\lambda^{\Delta_++1}{\rho_x}_{+}.
\end{equation}
We will use the transformation to fix the chemical potential for each solution
the same, i.e., we work in grand canonical ensemble.

Note that the set of equations admits an analytical solution with
vanishing $\rho_\mu$, corresponding to the normal phase (conductor
phase). This solution is just the AdS Reissner-Nordstr\"om black
hole, given by
\begin{equation}\label{RN}
\begin{split}
ds^2=-f(r)dt^2+\frac{dr^2}{f(r)}+r^2(dx^2+dy^2),\\
f(r)=r^2-\frac{1}{r}(r_h^3+\frac{\mu^2 r_h}{4})+\frac{\mu^2 r_h^2}{4r^2}, \quad \phi(r)=\mu(1-\frac{r_h}{r}),
\end{split}
\end{equation}
with the temperature $T=\frac{r_h}{4\pi}(3-\frac{\mu^2}{4r_h^2})$ and the entropy $S=\frac{2\pi V_2}{\kappa^2}r_h^2$.
\section{Free energy and dual stress-energy tensor}
\label{sect:free}
In order to determine which phase is thermodynamically favored, we should calculate the free energy of the system for both normal phase and condensed phase. We will work in grand canonical ensemble in this paper, where the chemical potential is fixed. In gauge/gravity duality the grand potential $\Omega$ of the boundary thermal state is identified with temperature $T$ times the on-shell bulk action in Euclidean signature. The Euclidean action must include the Gibbons-Hawking boundary term for a well-defined Dirichlet variational principle and further a surface counterterm for removing divergence. Since we consider a stationary problem, the Euclidean action is related to the Minkowski one by a minus sign as
\begin{equation}\label{onshell}
-2\kappa^2 S_{Euclidean}=\int dx^4\sqrt{-g}(\mathcal{R}+6+\mathcal{L}_m)+\int_{r\rightarrow\infty} d^3x\sqrt{-\bar{h}}(2\mathcal{K}-4),
\end{equation}
where $\bar{h}$ is the determinant of the induced metric $\bar{h}_{\mu\nu}$ on the boundary, and $\mathcal{K}$ is the trace of the extrinsic curvature $\mathcal{K}_{\mu\nu}$.~\footnote{In principle, we should also consider the surface counterterm for the charged vector field $\rho_\mu$, but one can easily see that this term makes no contribution under the source free condition, i.e., $\rho_{x-}=0$.}

Employing the equations of motion, the on-shell action reduces to
\begin{equation}\label{onshell_ext}
-2\kappa^2 S_{Euclidean}^{on-shell}=2\beta V_2 e^{-\chi/2}r\sqrt{fh}(\mathcal{K}r-2r-\sqrt{f})|_{r\rightarrow\infty},
\end{equation}
with $\beta=1/T$ and $V_2=\int dx dy$. Substituting the asymptotical expansion~\eqref{boundary} into~\eqref{onshell}, we obtain
\begin{equation}\label{free}
\Omega=T S_{Euclidean}^{on-shell}=\frac{V_2}{2\kappa^2}f_3,
\end{equation}
where we have used the condition $h_3=\chi_3$ which can be easily found form the equations of motion~\eqref{eoms}. Note that for the normal phase shown in~\eqref{RN}, one has $f_3=-r_h^3-\frac{\mu^2 r_h}{4}$, and $h_3=\chi_3=0$.

According to the AdS/CFT dictionary,
the stress-energy tensor of the dual field theory can be calculated
by~\cite{Balasubramanian:1999re}
\begin{equation}\label{stress}
T_{ij}=\frac{1}{\kappa^2}\lim_{r\rightarrow\infty}[r(\mathcal{K}\bar{h}_{ij}-\mathcal{K}_{ij}-2\bar{h}_{ij})],
\end{equation}
with $i, j=\{t,x,y\}$. By using of the asymptotical expansion~\eqref{boundary}, we have
\begin{equation}\label{stresstensor}
\begin{split}
T_{tt}&=\frac{1}{2\kappa^2}(-2f_3+3h_3),\\
T_{xx}&=\frac{1}{2\kappa^2}(-f_3+3h_3),\\
T_{yy}&=\frac{1}{2\kappa^2}(-f_3),
\end{split}
\end{equation}
with vanishing non-diagonal components. For the normal phase with $h_3=\chi_3=0$, we find that $T_{xx}=T_{yy}$ and $\Omega/V_2=-T_{yy}$; the former shows the isotropy in $x-y$ plane and the latter gives the correct thermodynamical relation for the dual field theory to the
 AdS Reissner-Nordstr\"om black hole. In the condensed phase with nonzero $\langle\hat{J_x}\rangle$, the rotational symmetry is broken, thus it is expected to have $T_{xx}\neq T_{yy}$.  But in both cases, the stress energy tensor is traceless, which is consistent with the fact that we are considering a dual conformal field theory at the AdS boundary.
\section{Phase transition}
\label{sect:superconducor}
In what follows we will look for condensed phases numerically. We take
different $m^2$'s into consideration and for each $m^2$ we scan a wide range of $q$ which determines the
strength of the back reaction of matter fields on the background. Our numerical results reveal that the
system exhibits distinguished behavior depending on concrete value of $m^2$. There exists a particular value of $m^2$,
for which we denote as $m_c^2$. In the case with $m^2>m_c^2$, the condensed phase seems to survive even down to sufficiently
low temperatures, i.e., $T\rightarrow0$. In contrast, in the case with $m^2<m_c^2$, the condensed phase cannot
 exist below a finite temperature. To determine the precise value for $m_c^2$, we need to solve the coupled equations of
 motion~\eqref{eoms} at very low temperatures to see whether the condensate would turn back to a higher temperature.
 The $T\rightarrow0$ limit is a challenge in numerical calculation. Nevertheless, our numerical calculation
 suggests that $m_c^2=0$, for which we have some to say below. We will consider one concrete example for both cases. In each case we find similar results for other values of $m^2$.

\subsection{$m^2=3/4$}
\label{sect:positive}
For the case with $m^2>m_c^2$, we choose $m^2=3/4$ as a concrete example. For each value of $q$, the AdS Reissner-Nordstr\"om solution always exists even down to the zero temperature limit.
However, for sufficiently low temperature, we always find additional solutions with non-vanishing $\rho_x$ that are thermodynamically preferred. That is to say, for each value of $q$ we take, there is a phase transition occurring at a certain temperature $T_c$, where a charged black hole developing vector hair becomes thermodynamically favored. In the dual field theory side, it means that a vector operator acquires a vacuum expectation value $\langle\hat{J_x}\rangle\neq0$ breaking the U(1) symmetry spontaneously. Furthermore, the condensate $\langle\hat{J_x}\rangle$ chooses a special direction, so the rotational symmetry in $x-y$ plane is also destroyed. Our numerical calculation indicates that the order of the phase transition can be changed from second order to first order as one increases the strength of the back reaction. More precisely, the phase transition is second order for $q>q_c$ and first order for $q<q_c$, where $q_c\simeq1.3575$ for $m^2=3/4$.
\begin{figure}[h]
\centering
\includegraphics[scale=1]{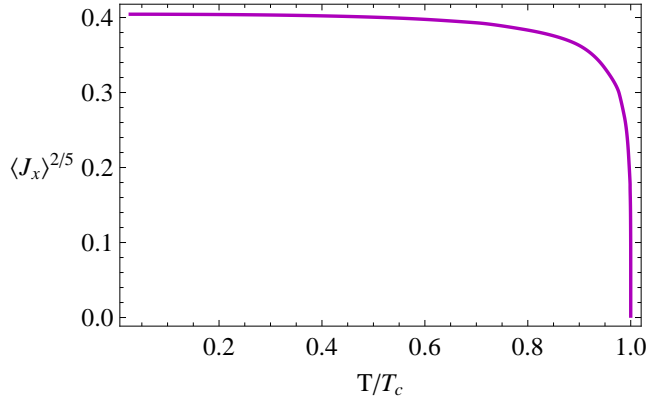}
\caption{\label{condensatepa} The condensate $\langle\hat{J_x}\rangle$ as a function of temperature. We choose $q=1.5$ and $m^2=3/4$. The condensate begins to appear at $T_c\simeq0.0179\mu$ and rises continuously as one further lowers the temperature, signaling a second order transition.}
\end{figure}
\begin{figure}[h]
\centering
\includegraphics[scale=0.95]{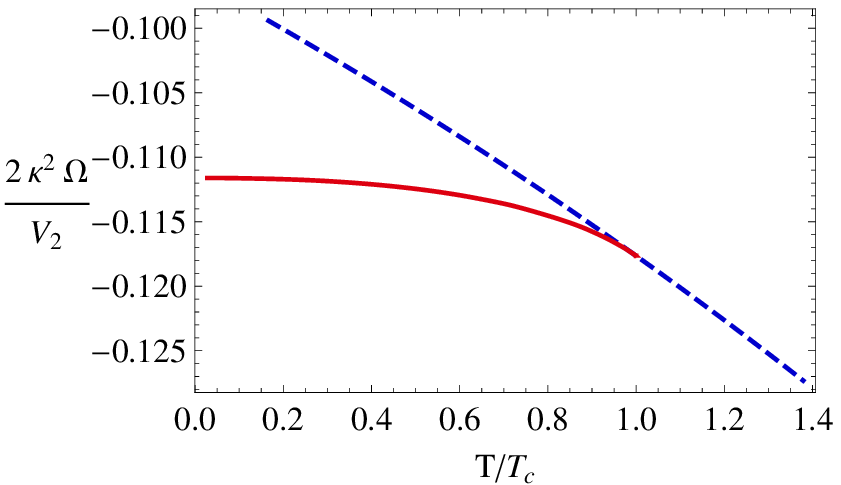}\ \ \ \
\includegraphics[scale=0.90]{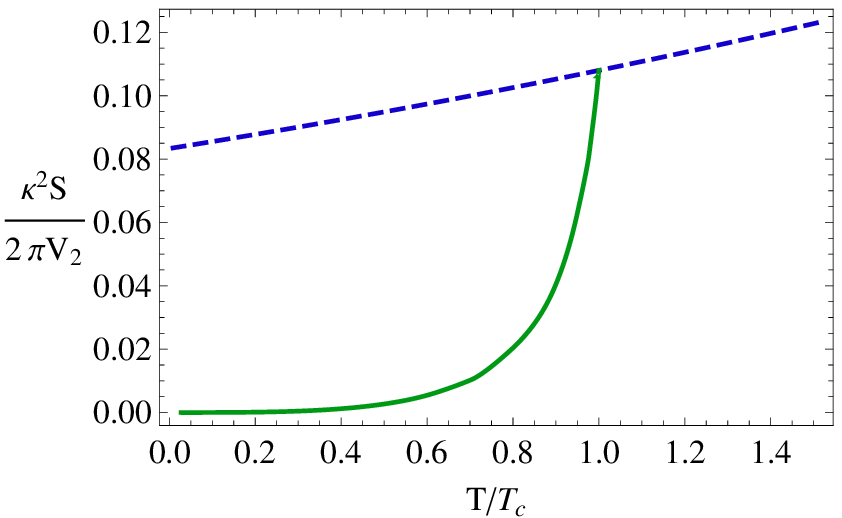}
 \caption{\label{freepya} The grand potential $\Omega$ (left plot) and
  thermal entropy $S$ (right plot) as a function of temperature. In both plots,
   the dashed blue curves are for the normal phase, while the solid curves are for
    the condensed phase. For $T>T_c$, one can only get the blue curve, but for lower
     temperature $T<T_c$ the condensed phase appears and has the lower free energy,
      thus is thermodynamically favored.}
\end{figure}

Taking $q=1.5>q_c$ as a typical example, apart from the AdS
Reissner-Nordstr\"om solution,  we find another set of solutions
with nonzero $\langle\hat{J_x}\rangle$ appearing below the critical
temperature $T_c$. Figure~\ref{condensatepa} presents the condensate
as a function of temperature, from which one can see that
$\langle\hat{J_x}\rangle$ rises continuously from zero at $T_c$. The
grand potential $\Omega$ is drawn in the left plot of
figure~\ref{freepya}. It is clear that below the critical
temperature $T_c$, the state with non-vanishing vector ``hair" is
indeed thermodynamically favored over the normal phase. We draw the
thermal entropy $S$ with respect to temperature in the right plot of
figure~\ref{freepya}. One can see that at the critical temperature
$T_c$, the entropy $S$ is continuous but its derivative has a jump,
indicating a second order phase transition. Our numerical results
also suggest that the critical exponent for all $q>q_c$ is always
$1/2$, i.e., $\langle\hat{J_x}\rangle\sim(1-T/T_c)^{1/2}$.
\begin{figure}[h]
\centering
\includegraphics[scale=1]{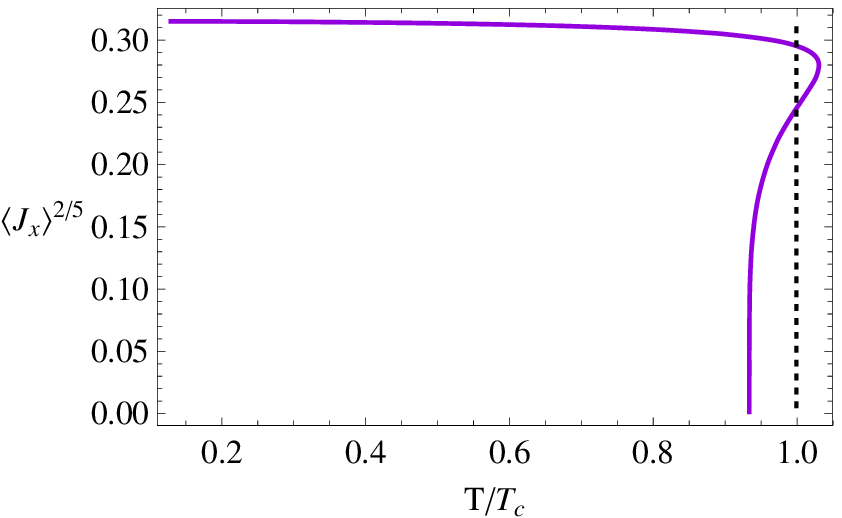}
\caption{\label{condensatepb} The condensate
$\langle\hat{J_x}\rangle$ as a  function of temperature for $q=1.2$
and $m^2=3/4$. The critical temperature $T_c\simeq0.00342\mu$ is
denoted by a vertical dotted line. The condensate becomes
multi-valued at $T\simeq1.03T_c$. The value of condensate has a
sudden jump from zero to the upper part of the purple solid curve at
$T_c$, indicating a first order transition.}
\end{figure}
\begin{figure}[h]
\centering
\includegraphics[scale=0.94]{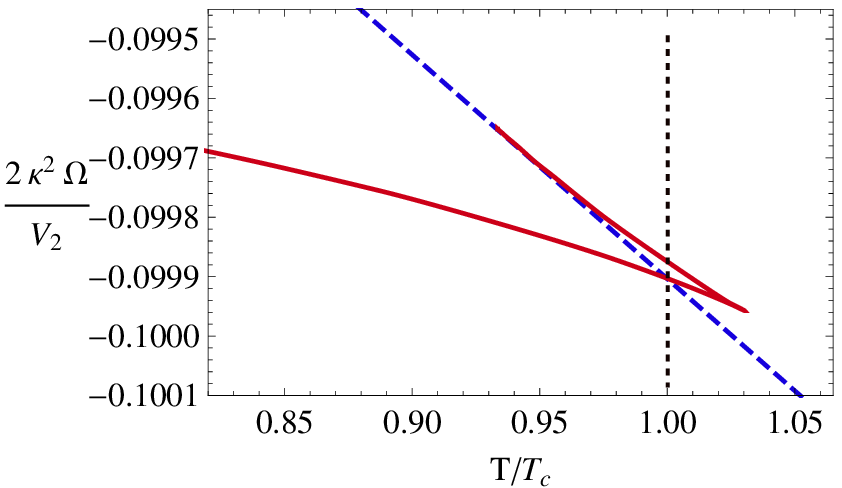}\ \ \ \
\includegraphics[scale=0.91]{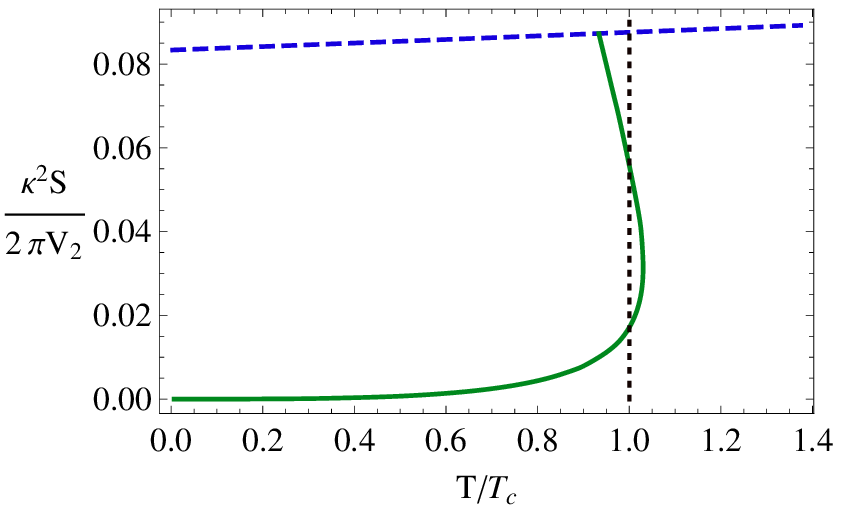}
\caption{\label{freepyb} The grand potential $\Omega$ (left plot)
and thermal  entropy $S$ (right plot) with respect to temperature
with $q=1.2$ and $m^2=3/4$. In both plots, dashed blue curves come
from the normal phase, while the solid curves come from the
condensed phase. Trace the physical curve by choosing the lowest
grand potential at a fixed $T$. The critical temperature at which
the condensed phase begins to be thermodynamically favored is
$T_c\simeq0.00342\mu$, denoted by a vertical dotted line. The
entropy jumps from the blue curve to the lowest branch of the green
solid curve at $T_c$.}
\end{figure}

A qualitative change happens as we decrease $q$ past $q_c$. Consider
the case  with $q=1.2<q_c$. The condensate $\langle\hat{J_x}\rangle$
versus temperature is presented in figure~\ref{condensatepb}.
Compared to the previous case, the condensate becomes multi-valued
and we can find two new sets of solutions with non-vanishing
$\langle\hat{J_x}\rangle$ at temperatures lower than
$T\simeq1.03T_c$, involving an upper-branch with large
$\langle\hat{J_x}\rangle$ and a down-branch with small
$\langle\hat{J_x}\rangle$. Therefore, there are three states that
are available to the system at some temperature, i.e., one is for
$\langle\hat{J_x}\rangle=0$ and two for
$\langle\hat{J_x}\rangle\neq0$. To determine which is the physical
state, we draw the grand potential $\Omega$ in figure~\ref{freepyb}.
One can find that the free energy versus temperature develops a
characteristic ``swallow tail" which is typical in first order phase
transition. The normal phase is thermodynamically favored at higher
temperatures, but as we lower the temperature down to $T_c$, the
upper-branch finally dominates the system. We present the entropy
$S$ versus temperature in the right part of figure~\ref{freepyb},
from which one can see that $S$ is also multi-valued and has a
sudden jump from the normal phase to the physical condensed phase at
$T_c$. Clearly, the transition is first order.
\begin{figure}[h]
\centering
\includegraphics[scale=1]{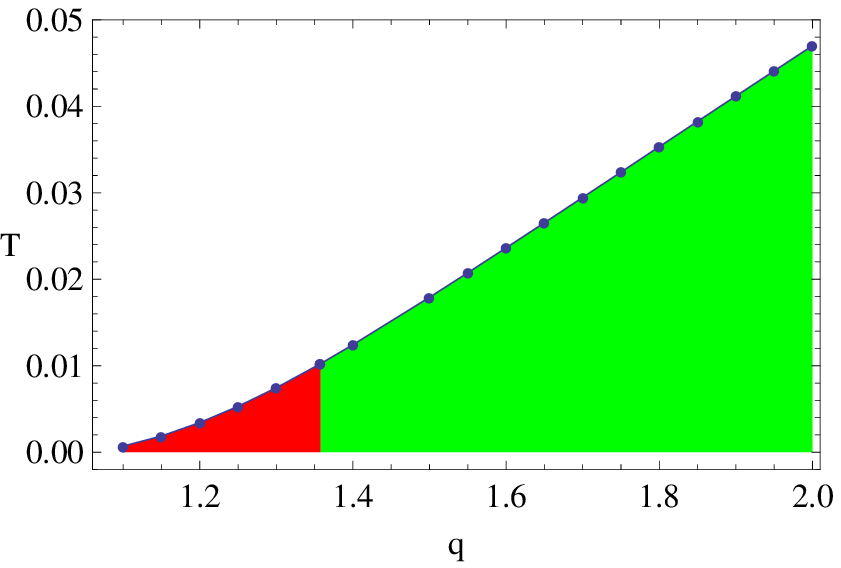}
\caption{\label{phasediagp} The phase diagram for $m^2=3/4$. The
solid curve  separates the condensed phase from the normal phase.
The critical value $q_c$ divides the condensed phase into two parts.
The case $q>q_c$ is associated with second order phase transition
(green area), while $q<q_c$ corresponds to first order transition
(red area).}
\end{figure}

One interesting feature presented in both cases is that the hairy
black hole  exhibits tiny entropy at finite low temperatures,
compared with the normal phase in figure~\ref{freepya} and
figure~\ref{freepyb}. Since the values of $S$ are obtained from the
behavior of the solutions at the horizon, it is difficult to extract
them with high accuracy at sufficiently low temperature.
Nevertheless, our numerical results suggest that entropy remains
small and smoothly decreases as the temperature is gradually
lowered. As being a single state without any degeneracy, a
superconducting ground state should not have any entropy. In the
gravity side, it corresponds to the fact that the zero temperature
limit of the superconducting black holes should have zero horizon
area, which was previously observed in
refs.~\cite{Gubser:2009gp,Horowitz:2009ij}.

The main results of this subsection are summarized by the ($T$, $q$)
phase diagram shown  in figure~\ref{phasediagp}. The solid curve
gives the critical temperature $T_c$ for the phase transition from
the normal phase to the condensed phase. There is a critical value
of $q$, denoted as $q_c$, above which the phase transition is second
order, while below which the transition becomes first order. It is
also clearly that as $q$ decreases, $T_c$ decreases gradually, which
tells us that the increase of the back reaction hinders the phase
transition.
\subsection{$m^2=-3/16$}
\label{sect:negative}
Similar to the previous discussion, for each $m^2<m_c^2$ we scan a wide range of $q$ to find all possible types of transitions.
There exist two special values of $q$, denoted as $q_\alpha$ and $q_\beta$ with $q_\alpha>q_\beta$, which divides the
parameter space of $q$ into three regions, $q>q_\alpha$, $q_\beta<q<q_\alpha$ and $q<q_\beta$, respectively.
The thermodynamic behavior changes qualitatively in three regions. We may find second order transition, first order
 transition and zeroth order transition. In order to take account of the order of the phase transitions,
 we denote the transition temperature as $T_2$, $T_1$ and $T_0$, respectively. As a typical example, we will
 take the mass parameter $m^2=-3/16$ in this subsection. In this case $q_\alpha\simeq 1.0175$ and $q_\beta\simeq 0.9537$.
 Details are given as follows.
\begin{figure}[h]
\centering
\includegraphics[scale=0.90]{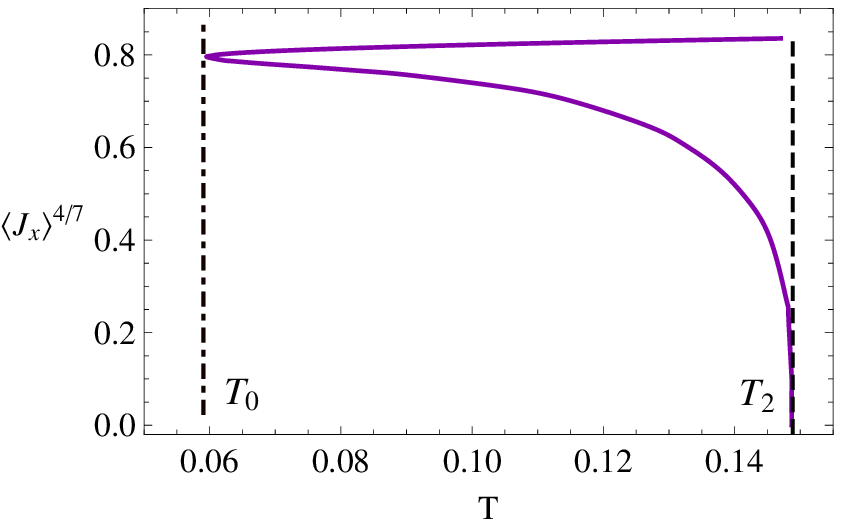}\ \ \ \
\includegraphics[scale=0.93]{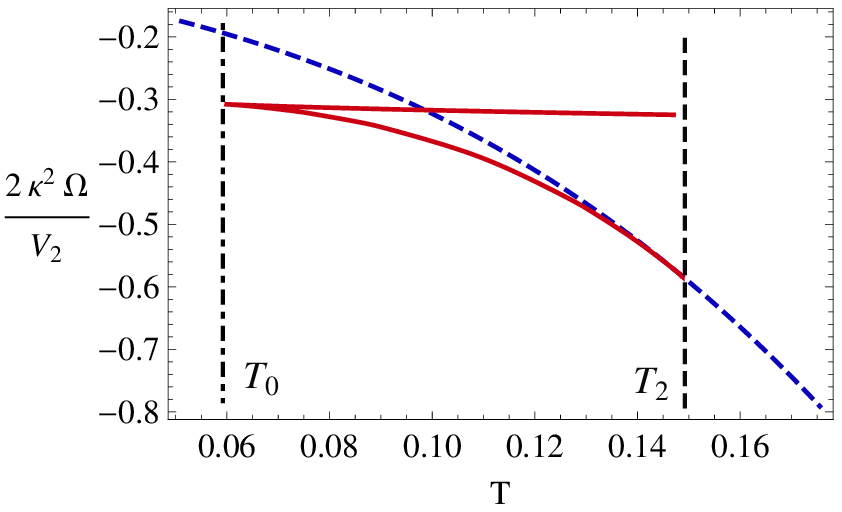}
\caption{\label{condfreea} The condensate $\langle\hat{J_x}\rangle$
(left) and free energy $\Omega$ (right) as a function of temperature
for $q=2$ and $m^2=-3/16$. The dashed blue curve is from the normal
phase, while the solid curves are from the condensed phase.
$T_2\simeq0.1487\mu$ is denoted by a vertical dashed line and
$T_0\simeq0.05972\mu$ is denoted by a vertical dot dashed line. The
condensate becomes multi-valued between $T_0$ and $T_2$. The lower
branch of the condensed phase has the lowest free energy. For
temperatures $T<T_0$ and $T>T_2$, the normal phase is
thermodynamically preferred.}
\end{figure}
\begin{figure}[h]
\centering
\includegraphics[scale=0.89]{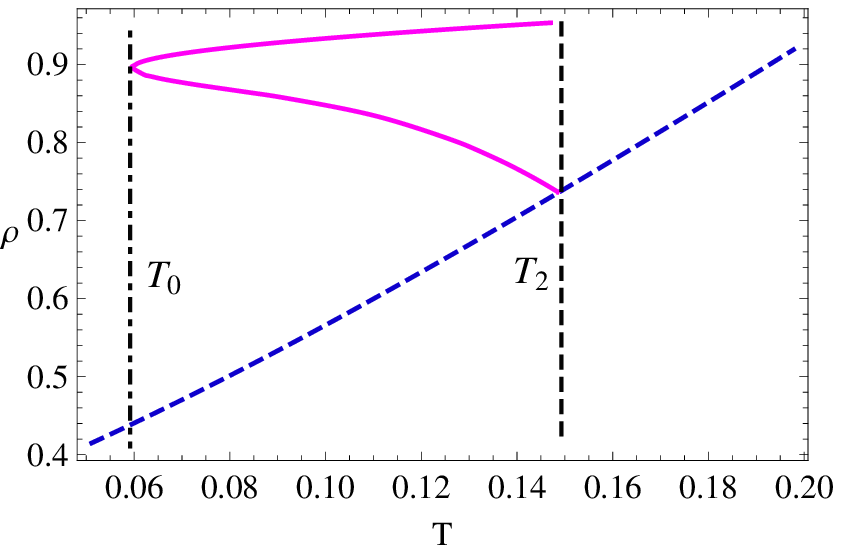}\ \ \ \
\includegraphics[scale=0.96]{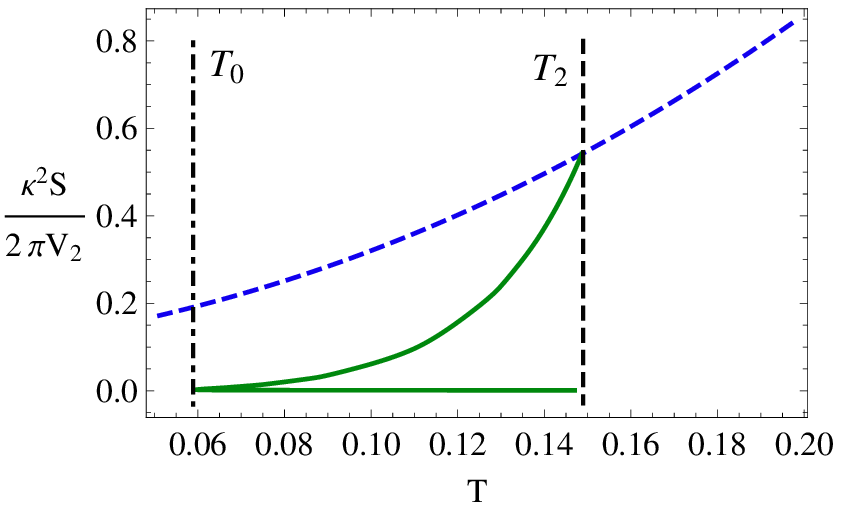}
\caption{\label{chargentropya} The charge density $\rho$ (left) and
thermal entropy $S$ (right) versus temperature for $q=2$ and
$m^2=-3/16$. The dashed blue curves are for the normal phase, while
the solid curves for the condensed phase. $T_2\simeq0.1487\mu$ is
denoted by a vertical dashed line and $T_0\simeq0.0597\mu$ is
denoted by a vertical dot dashed line. As $T_0<T<T_2$, $\rho$ and
$S$ become multi-valued between, where only the lower branch of the
charge density and the upper branch of the entropy are
thermodynamically preferred. In other region of temperature, the
normal phase dominates the phase diagram. Both $\rho$ and $S$ are
continuous but not differentiable at $T_2$, characterizing a second
order phase transition.}
\end{figure}

For the case with small back reaction, i.e., $q>q_\alpha$, we focus
on the case with $q=2$.  The condensate versus temperature is
exhibited in the left plot of figure~\ref{condfreea}. We immediately
see that $\langle\hat{J_x}\rangle$ is multi-valued above the
temperature denoted as $T_0\simeq0.05972\mu$. Similar to the first
order transition for $m^2=3/4$, the condensed phase has two
branches, i.e., the upper-branch with large
$\langle\hat{J_x}\rangle$ and a down-branch with small
$\langle\hat{J_x}\rangle$. The free energy $\Omega$ drawn in the
right plot of figure~\ref{condfreea} also shows a ``swallow tail"
shape, but it is very different from the one in
figure~\ref{freepyb}. Comparing the free energy $\Omega$ for each
solution, we find that the condensed solutions in the down-branch
are thermodynamically favored, which only exist in a small range
$T_0<T<T_2$. At other temperatures, including $T<T_0$ and $T>T_2$,
it is the normal phase with $\langle\hat{J_x}\rangle=0$ that is
thermodynamically relevant. At the temperature $T_0$, the free
energy has a sudden jump from the condensed phase to the normal
phase, indicating a zeroth order transition. The charge density
$\rho$ and entropy $S$ in figure~\ref{chargentropya} show that both
$\rho$ and $S$ are continuous but have a kink at $T_2$, indicating a
second order transition. It is interesting to note that the
condensed phase terminates at a finite lower temperature $T_0$.
\begin{figure}[h]
\centering
\includegraphics[scale=0.885]{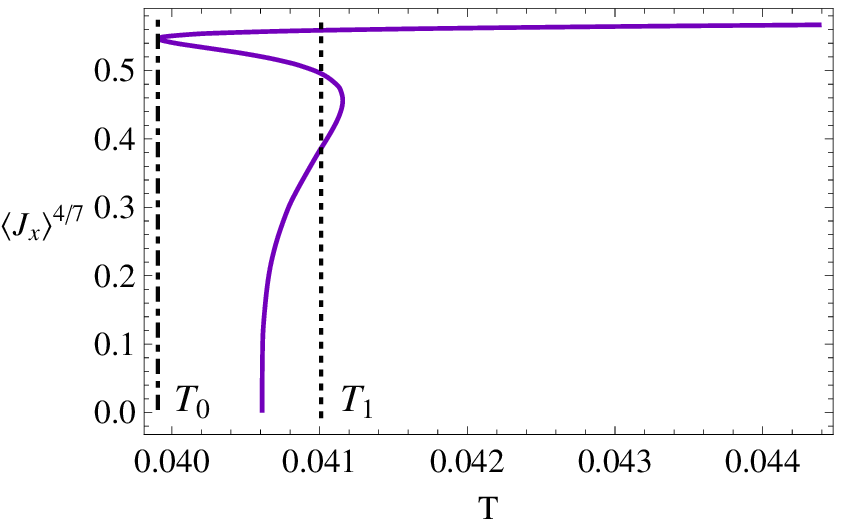}\ \ \ \
\includegraphics[scale=0.965]{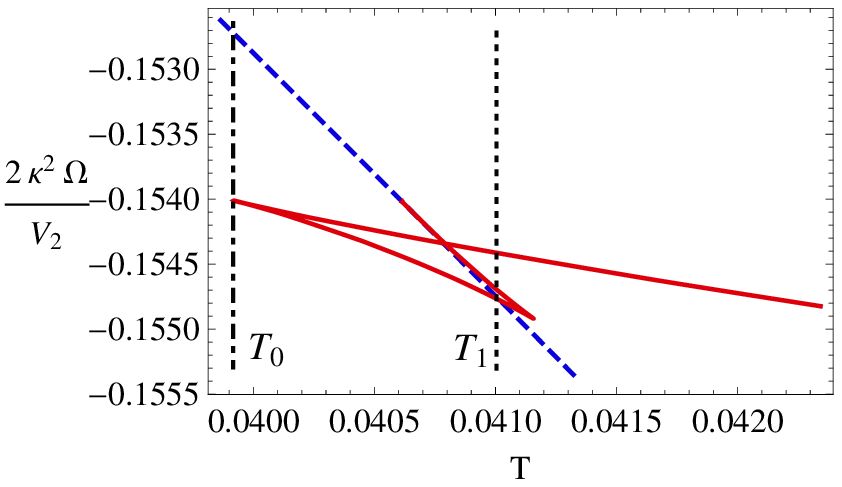}\caption{\label{condfreeb}
The condensate $\langle\hat{J_x}\rangle$ (left) and free energy $\Omega$ (right) as a function
of temperature for $q=39/40$ and $m^2=-3/16$. The dashed blue curve is from the normal phase,
while the solid curves are from the condensed phase. $T_1\simeq0.04102\mu$ is denoted as a
vertical dashed line and $T_0\simeq0.03992\mu$ is denoted as a vertical dot dashed line.
The condensate behaves multi-valued. The middle branch of the condensed phase has the
lowest free energy between $T_0$ and $T_1$. For other range of temperature, the normal phase is thermodynamically favored.}
\end{figure}
\begin{figure}[h]
\centering
\includegraphics[scale=0.89]{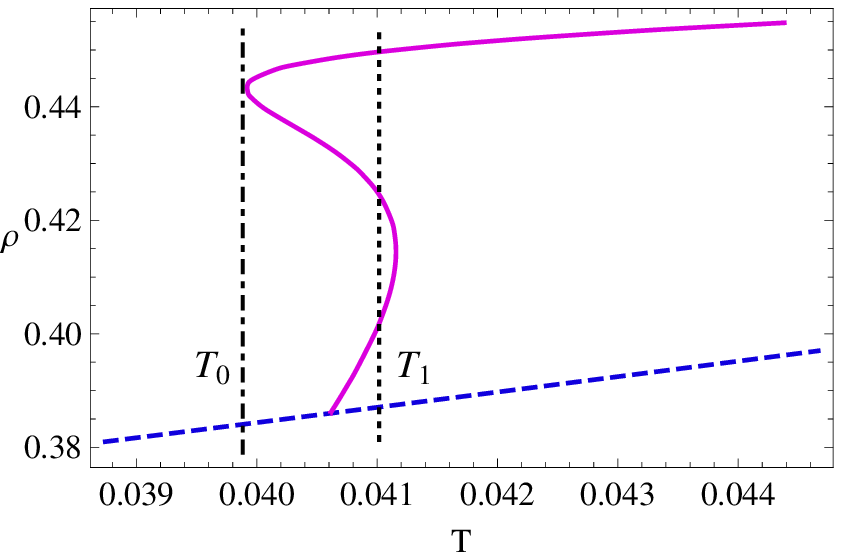}\ \ \ \
\includegraphics[scale=0.96]{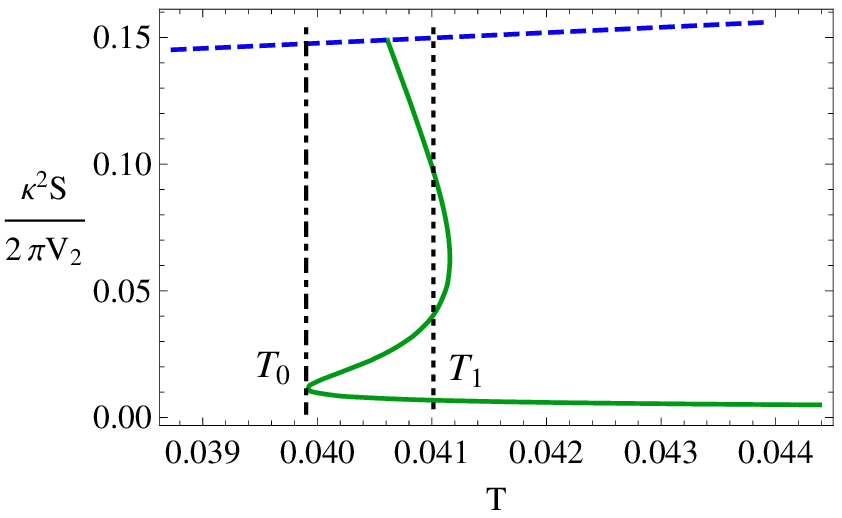}
\caption{\label{chargentropyb} The charge density $\rho$ (left) and
thermal entropy $S$ (right) versus temperature for $q=39/40$ and
$m^2=-3/16$.  The dashed blue curves correspond to the normal phase,
while the solid curves to the condensed phase. $T_1\simeq0.04102\mu$
is denoted by a vertical dotted line and $T_0\simeq0.03992\mu$ is
denoted by a vertical dot dashed line. When $T_0<T<T_1$, $\rho$ and
$S$ are both multi-valued. In both plots, it is the middle branch
that is thermodynamically preferred. In other region of temperature,
the normal phase is thermodynamically favored. Both $\rho$ and $S$
are not continuous at $T_1$, but rather jump from the blue dashed
line to the middle branch of the solid line, signaling a first order
transition.}
\end{figure}

For the case $q_\beta<q<q_\alpha$, let us consider for example the case with $q=39/40$. The behaviors of condensate and other thermodynamical quantities are much more complicated. Figure~\ref{condfreeb} plots the condensate with respect to temperature, where $\langle\hat{J_x}\rangle$ is also multi-valued above $T_0\simeq0.03992\mu$. But there are three sets of condensed solutions. According to the value of condensate, we denote them as the upper-branch for large $\langle\hat{J_x}\rangle$, the middle-branch for middle $\langle\hat{J_x}\rangle$ and the down-branch for small $\langle\hat{J_x}\rangle$, respectively. We also draw the grand potential $\Omega$ in the right plot of figure~\ref{condfreeb}. We present the charge density as well as the thermal entropy in figure~\ref{chargentropyb}. The values of $\rho$ and $S$ have a sudden jump from the normal phase to the thermodynamically favored branch of the condensed phase at $T_1$, indicating a first order phase transition. As we lower the temperature, the phase with $\langle\hat{J_x}\rangle=0$ is first thermodynamically favored, and then the middle-branch begins to dominate the thermodynamics through a first order transition at $T_1\simeq0.04102\mu$, finally the condensed phase ends up at the temperature $T_0$ where a zeroth order transition appears. Our numerical results uncover that as one increases the strength of the back reaction, $T_1$ decreases while $T_0$ also decrease but with a slowly rate, and finally $T_1$ becomes equal to $T_0$ at $q_\beta$. For the present case $q_\beta<q<q_\alpha$, the value of $T_1$ is always larger than $T_0$, so we have a first order transition from the normal phase to the condensed phase at higher temperature $T_1$ and then a zeroth order transition from the condensed phase to the normal phase at lower temperature $T_0$.
\begin{figure}[h]
\centering
\includegraphics[scale=0.92]{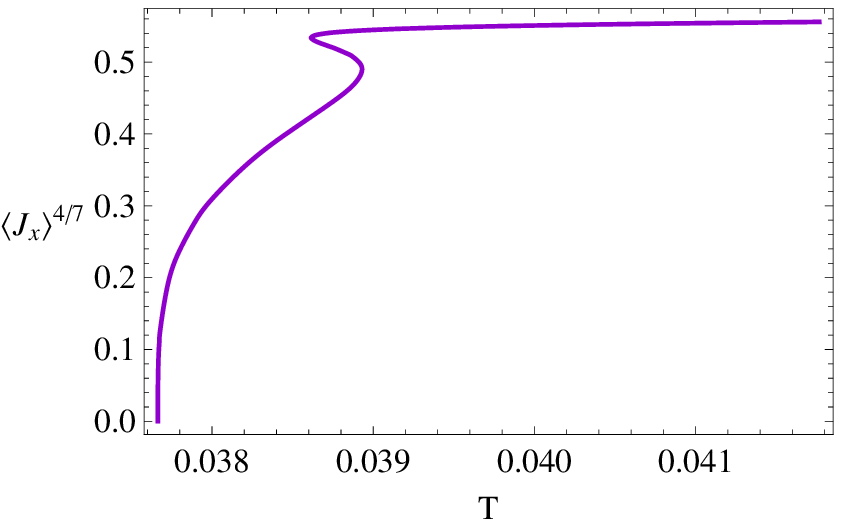}\ \ \ \
\includegraphics[scale=0.92]{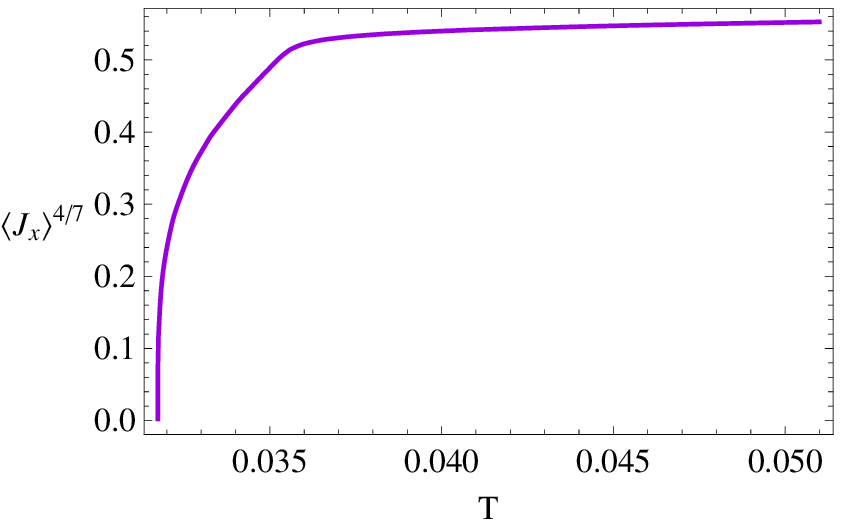}
\caption{\label{condensatec} The condensate $\langle\hat{J_x}\rangle$ as a function of temperature for $q=19/20$ (left) and $q=9/10$ (right). The condensate only emerges above the temperature $T_n\simeq0.03766\mu$ for $q=19/20$ and $T_n\simeq0.03174\mu$ for $q=9/10$. The condensate in the left plot behaves multi-valued.}
\end{figure}

For the above two examples, an interesting common feature is that the thermodynamically
favored hairy black hole solutions exist up to a minimal temperature $T_0$, where it connects
 with an unstable condensed branch starting from higher temperature, which will be discussed below.
 In the present model we restrict the case with $\rho_x$ turned on, only the uncondensed phase can
 appear below $T_0$, so the free energy has a sudden jump from the condensed phase to the normal phase at $T_0$,
  indicating a zeroth order transition. Note that in the theory of superfluidity and superconductivity, a
  discontinuity of the free energy was discovered theoretically and an exactly solvable model for
  such phase transition was given in ref.~\cite{Maslov:2004}. Therefore, it is quite interesting to
  see whether the holographic model has some relation to the model in ref.~\cite{Maslov:2004}.

\begin{figure}[h]
\centering
\includegraphics[scale=0.92]{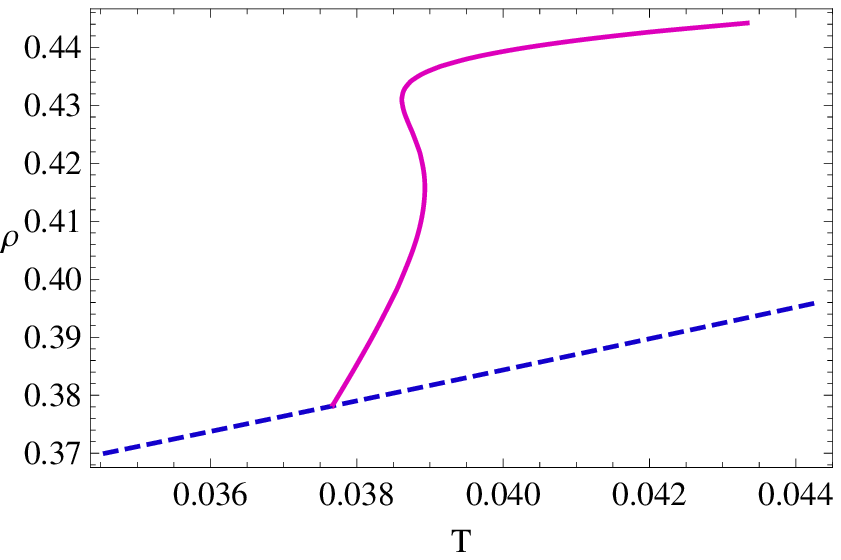}\ \ \ \
\includegraphics[scale=0.92]{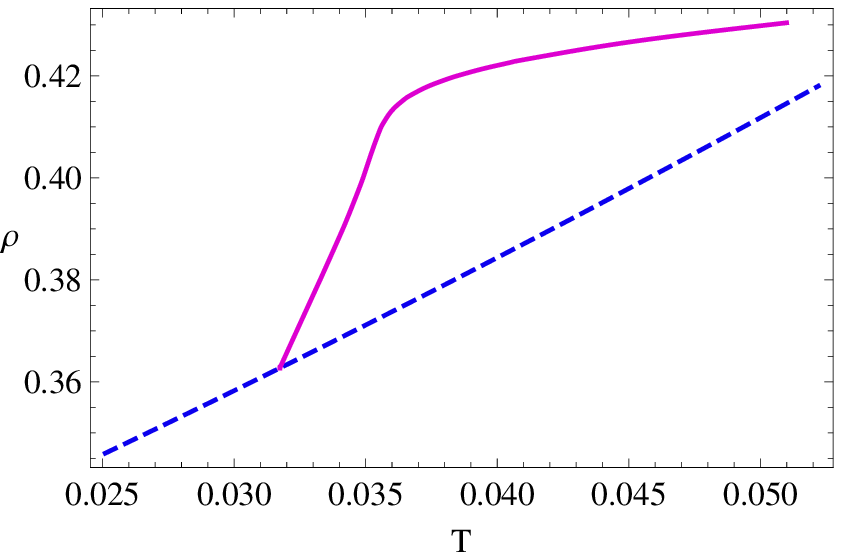}
\caption{\label{chargenc} The charge density $\rho$ as a function of temperature for $q=19/20$ (left) and $q=9/10$ (right). The dashed blue curves come from the normal phase, while the solid curves from the condensed phase.}
\end{figure}
\begin{figure}[h]
\centering
\includegraphics[scale=0.92]{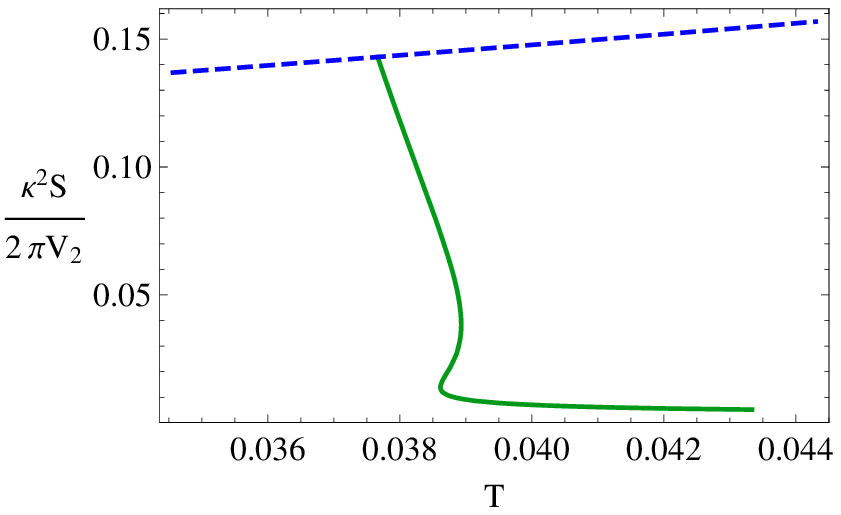}\ \ \ \
\includegraphics[scale=0.92]{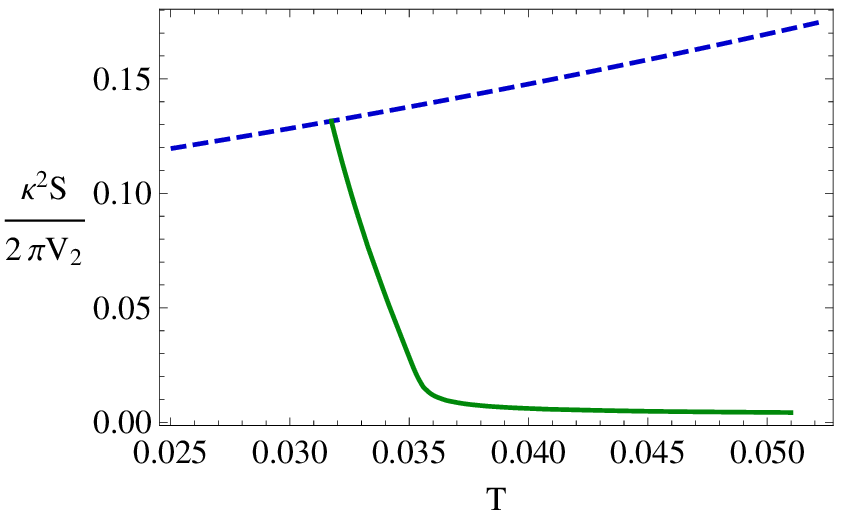}
\caption{\label{entropync} The thermal entropy $S$ as a function of temperature for $q=19/20$ (left) and $q=9/10$ (right). The dashed blue curves come from the normal phase, while the solid curves from the condensed phase.}
\end{figure}
\begin{figure}[h]
\centering
\includegraphics[scale=0.93]{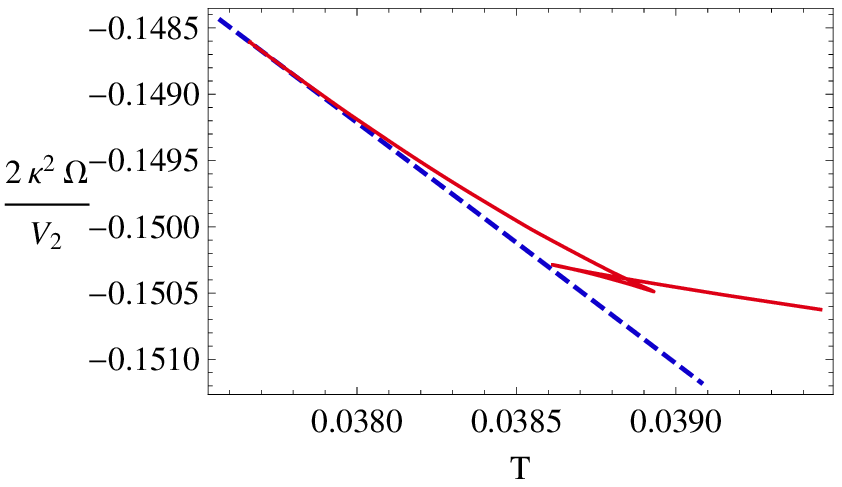}\ \ \ \
\includegraphics[scale=0.91]{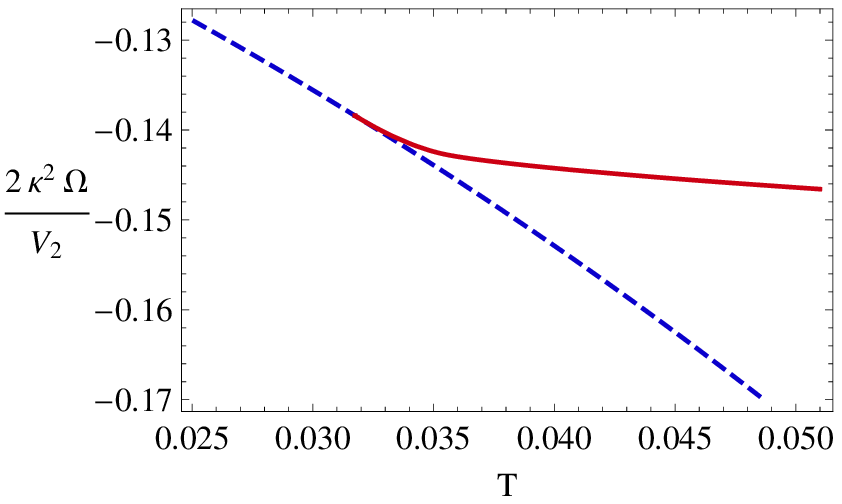}
\caption{\label{freenc} The free energy $\Omega$ as a function of
temperature for $q=19/20$ (left)  and $q=9/10$ (right). The dashed
blue curves come from the normal phase, while the solid red  curves
from the condensed phase. The free energy of the condensed phase in
the left plot forms a typical swallow tail. In both plots, the
condensed phase has the free energy larger than  the normal phase,
and thus is not thermodynamically preferred.}
\end{figure}

As $q$ decreases past $q_\beta$, we see a dramatic change in the
thermodynamics. We draw the condensate versus temperature in
figure~\ref{condensatec} for $q=19/20$ and $q=9/10$, from which we
can find that hairy solutions only appear at temperatures above
$T_n$ and the general trend is that $\langle\hat{J_x}\rangle$
increases with the temperature.~\footnote{One may wonder if the
curve could turn back at a higher temperature. We numerically check
this increasing trend up to a very high temperature as we can and
find the curve  has a well defined asymptotic behavior. Therefore,
we believe it will not turn around.} The value of
$\langle\hat{J_x}\rangle$ is multi-valued for the case with larger
$q$. As we decrease $q$, this multi valuedness disappears. For
completeness, we show the charged density $\rho$ in
figure~\ref{chargenc} and the thermal entropy $S$ in
figure~\ref{entropync}. Similar feature can also be found in these
two figures. At a first glance, this appears to be surprising, since
in  general one expects the condensed phase to emerge at low
temperatures rather than at high temperatures. To have a physical
condensed phase, the hairy black hole configuration should have free
energy less than the AdS Reissner-Nordstr\"om black hole describing
the normalphase. Comparing $\Omega$ between the condensed phase and
normal phase, we can clearly see in figure~\ref{freenc} that the
condensed phase has free energy much larger than the normal phase
and thus is not thermodynamically favored. Therefore these hairy
black holes represent unstable branches that do not contribute to
the thermodynamics. Similar phenomenon was previously found in
ref.~\cite{Buchel:2009ge} through a phenomenological model, known as
``exotic hairy black holes". Such a phenomenon also exists in some
consistent truncations of string/M-theory in
refs.~\cite{Donos:2011ut,Aprile:2011uq} as well as inhomogeneous
black hole solutions in AdS space dual to spatially modulated phase
of a field theory at finite chemical
potential~\cite{Withers:2013loa}. This phenomenon of a
thermodynamically subdominant condensate at higher temperature is
known as ``retrograde condensation".~\footnote{The terminology
``retrograde condensation" was first introduced to describe the
behavior of a binary mixture during isothermal compression above the
critical temperature of the mixture~\cite{Kuenen:1892}. A
subdominant condensate can exist in this system in some temperature
range.}
\begin{figure}[h]
\centering
\includegraphics[scale=0.7]{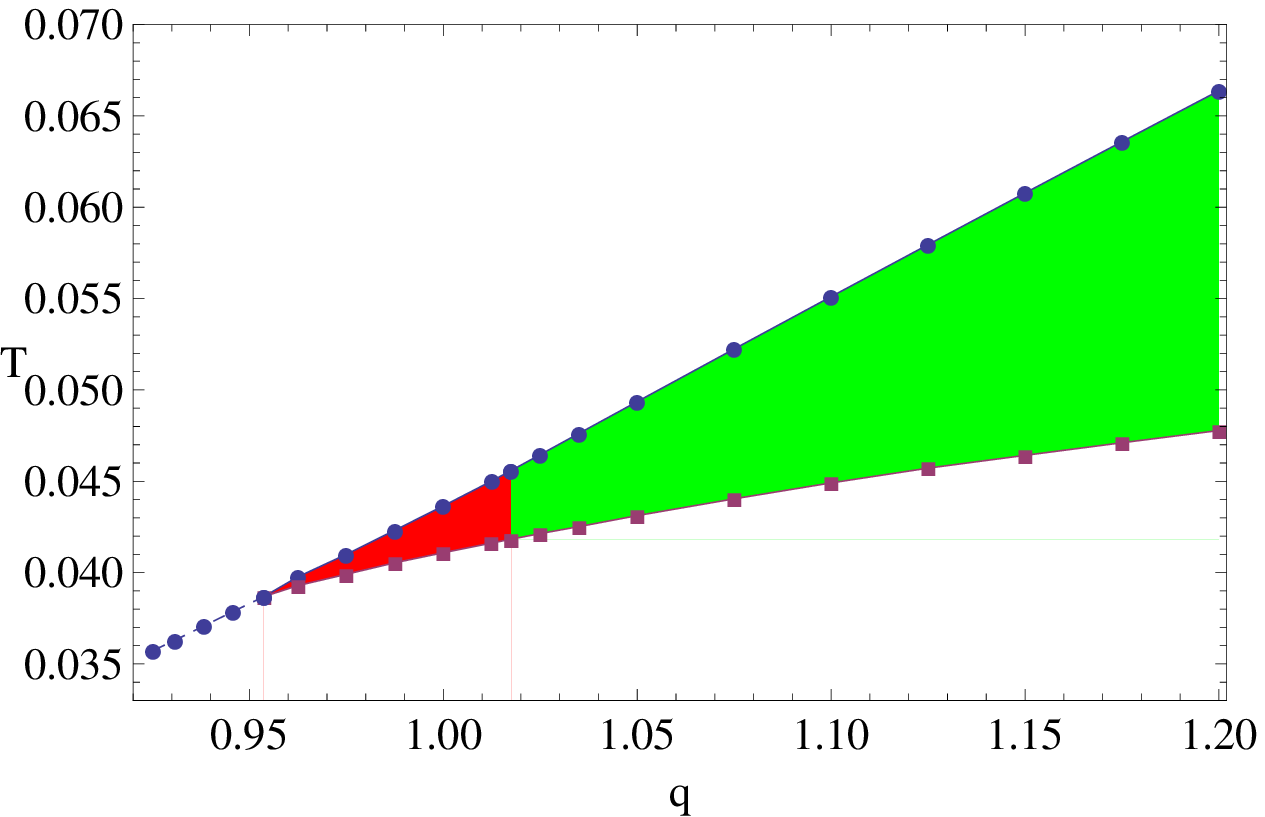}
\caption{\label{phasediagn} The ($T$, $q$) phase diagram for $m^2=-3/16$. The green region is related to the condensed phase through a second order transition from the normal phase, while the red region is associated with the case by a first order transition. For other areas, the normal phase is thermodynamically favored. The transition temperature $T_n$ for the retrograde condensation is denoted by the dashed line in the left down corner.}
\end{figure}

We summarize the main results of this subsection by constructing the ($T$, $q$) phase diagram in figure~\ref{phasediagn}. The upper solid curve indicates the transition from the normal phase to the condensed phase as we lower the temperature, which is the combination of $T_2$ for $q>q_\alpha$ and $T_1$ for $q_\beta<q<q_\alpha$. The lower solid curve represents the minimal temperature $T_0$ at which hairy solutions terminate. The region between the two boundary curves shrinks as one increases the strength of the back reaction and the two curves intersect at
$q_\beta$. The dashed curve for $q< q_\beta$ gives the value of $T_n$, above which a thermodynamically subdominant condensed phase appears.
\begin{figure}[h]
\centering
\includegraphics[scale=0.7]{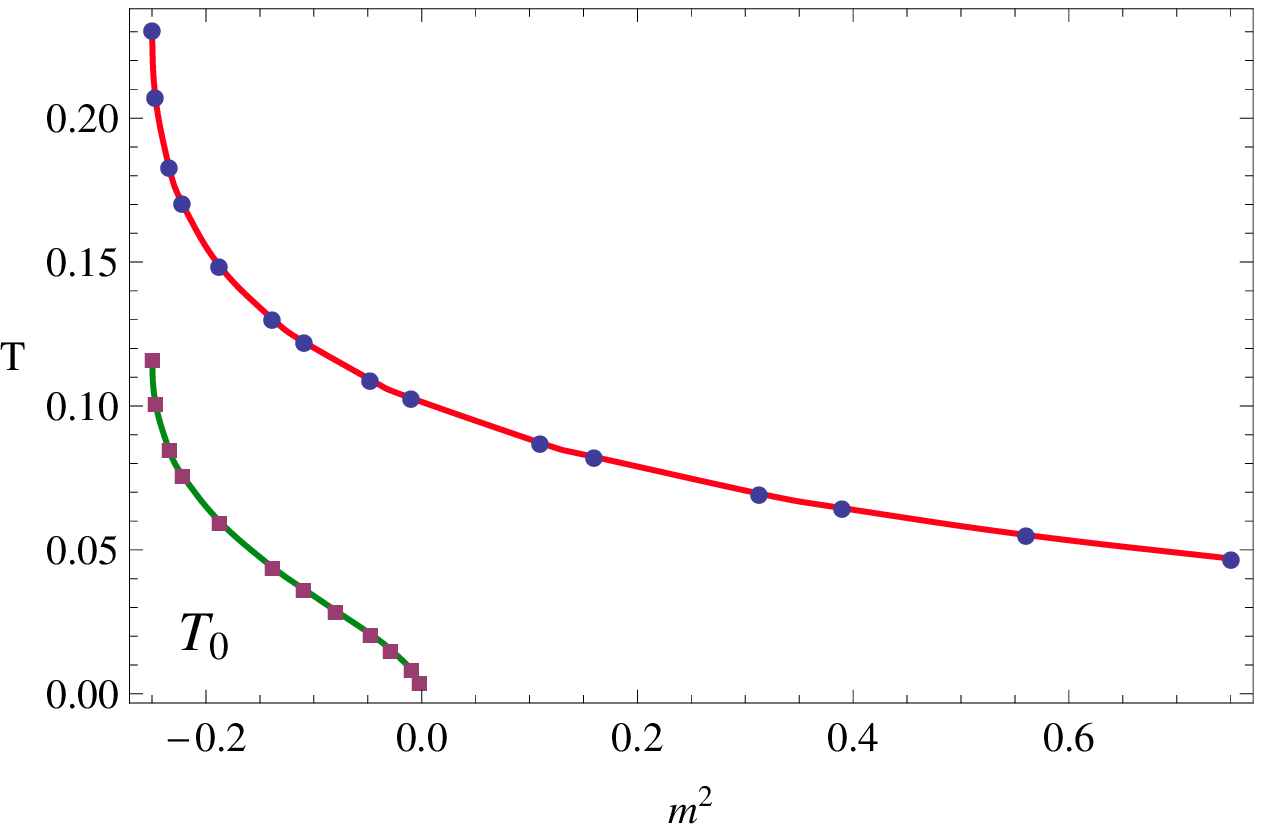}
\caption{\label{masstempe} The transition temperature from the normal phase to the condensed phase (upper curve) and the turning temperature $T_0$ (lower curve) with respect to $m^2$. Here $q=2$. The leftmost two points correspond to the mass square slightly above $m^2=-1/4$.}
\end{figure}

We now return to the critical mass $m_c^2$. For very small $m^2$, we
can find the temperature $T_0$ easily by directly numerical
calculation. Figure~\ref{masstempe} presents the temperature $T_0$
for each $m^2$. We find that the value of $T_0$ decreases quickly as
one increases $m^2$ from its lower bound. But, to search for $T_0$
numerically becomes more and more difficult as the value of $m^2$ is
close to the critical one. Due to the lake of numerical control at
sufficiently low temperatures, we are not able to give the values of
physical quantities, such as condensate $\langle\hat{J_x}\rangle$,
charge density $\rho$ and entropy $S$, at very low $T$.
Nevertheless, believing  $T_0$ as a function of $m^2$ exhibits a
well-behaved behavior and using the extrapolation, we find that the
value of $m_c^2$ locates at $m_c^2=0$ up to a numerical
error~\footnote{In the case with $m^2=-399/160000$ and $q=2$, we
find the turning point is $T_0\simeq0.00429\mu$. For the case with
$m^2=0$ and $q=2$, we numerically solve the model up to the
temperature $T\simeq0.00038\mu$ with very high computational
accuracy and find no turning point.}. There are also two other hints
that support our numerical result for the value of $m_c^2$: (1)
Comparing our model to the SU(2) p-wave model, the effective mass of
the vector operator is equal to zero for the latter,  we find that
the reduced equations of motion look very similar with each other
and the asymptotical expansions near the boundary are the same.
Since there does not exist any particular temperature at which the
condensate terminates and turns around to high temperatures in the
SU(2) p-wave case, it is natural to expect that the result is also
true in our model with $m^2=0$. Indeed, for the special case
$m^2=0$, if we ignore the possible turning point $T_0$ at very low
temperature, the model exhibits extremely similar behavior as the
SU(2) p-wave model~\cite{Ammon:2009xh,Akhavan:2010bf}. Furthermore,
our ongoing analysis of our model with the full back reaction in
(4+1) dimensional black hole  as well as soliton background cases
exhibits all known phase structure presented in this paper. (2)
Looking at the asymptotical expansion of $\rho_x$
in~\eqref{boundary}, one can see that the leading term with
coefficient $\rho_{x-}$ for $m^2>0$ is divergent as $ r \to \infty$.
In this case such a term must be regarded as source term and set to
be zero. In contrast, this term for $m^2<0$ is regular as
$r\rightarrow\infty$. Similar to the case in the s-wave
model~\cite{Hartnoll:2008vx}, we may have freedom to consider
$\rho_{x-}$ either as source term or expectation value of the dual
operator. The critical value of mass square is $m^2=0$. Thus, it is
not surprised that the model exhibits distinguished behaviors for
$m^2<0$ and $m^2>0$ numerically.

One may suspect that even for very large $m^2$ there exists a non
zero but very small minimal temperature like $T_0$. Indeed, we can
not rule out this possibility by numerical approach only.
Nonetheless, for large $m^2>m_c^2$, our numerical calculation
indicates that the physical branch of $\langle\hat{J_x}\rangle$
($\rho$ and $S$) versus temperature may behave well up to zero
temperature. In particular, we do not find any evidence of the
branch turning back to a one at higher temperature as a
characteristic of a zeroth order transition shown, for example, in
figure~\ref{condfreea}.~\footnote{As a typical example, we choose
the model parameters $m^2=3/4$ and $q=43/40$. We manage to solve the
coupled equations of motion~\eqref{eoms} up to the temperature as
low as $T\simeq3.332\times10^{-6}\mu$ and find no evidence of the
condensate turning around to a branch extending to the high
temperature region.} Of course it is helpful to clear up this issue
by constructing the extremal limit of the hairy black hole
solutions. We leave this for future investigation.

We draw the plot of the critical temperature versus the mass of the vector
field from the normal phase to the
condensed phase in figure~\ref{masstempe}. One can see that the
critical temperature decreases with the mass of $\rho_\mu$ at a
given $q$. This behavior is also observed in holographic models
involving a charged scalar field~\cite{Denef:2009tp,Gubser:2009qm}.

We did not find any hint in the probe limit that the condensate as
well  as other quantities turns back to a branch extending to the
high temperature region at a certain temperature like
$T_0$~\cite{Cai:2013pda}. Therefore, the appearance of $T_0$ is the
consequence of the back reaction. For sufficiently large $q$, the back
reaction can be ignored and we should recover the results in the
probe limit. In other words, $T_0$ should be zero and the critical
temperatures $T_c$ and $T_2$ should arrive at some finite constants as
$q\rightarrow\infty$. However, one can see from figure~\ref{phasediagn}
that the value of $T_0$ increases with $q$, and from figure~\ref{phasediagp} and
figure~\ref{phasediagn} that the critical temperatures $T_c$ and $T_2$ from
the normal phase to the condensed phase  also
increase with large $q$ linearly. This is seemingly inconsistent with the
expectation in the probe limit. In fact our numerical results are consistent with
the probe limit. The reason is as follows. Note that the probe limit demands that one
takes the limit $q\rightarrow\infty$, while keeping $q\rho_\mu$ and $q
A_\mu$ fixed in our model. To compare our results to the ones in the probe limit,
we should make the scaling transformation $\rho_\mu\rightarrow
q\rho_\mu$ and $A_\mu\rightarrow qA_\mu$. Under such a transformation,
the chemical potential $\mu$ becomes $q\mu$ and the temperature $T$
changes to $T/q$. Therefore, it is the value $T/q$ that corresponds to the temperature in
the probe limit. Indeed, in our numerical calculations, we checked
that $T_0/q \rightarrow 0$ and the critical temperatures $T_c/q$ and
$T_2/q$ approach to some constants as $q\rightarrow\infty$.  Therefore our
numerical results are in agreement with the probe limit analysis.

Furthermore, we know from from (\ref{stresstensor}) that
to see whether the dual stress-energy tensor is isotropic or not,
we have to extract the value of $h_3$ carefully in our calculations. We find
that the numerical solutions presented in our paper have
$h_3=0$ up to a numerical error ($\sim 10^{-14}$). It means that although the model has
an anisotropic structure associated with the p-wave order with
non-vanishing $\langle\hat{J_x}\rangle$ in the condensed phase, the
stress-energy tensor is isotropic. This  is the same as
 in the SU(2) p-wave model case~\footnote{The anisotropy of the
stress-energy tensor in the SU(2) model is controlled by the
constant $f_2^b$ appearing in equation (21) of
ref.~\cite{Ammon:2009xh}. It has been confirmed that the value of
$f_2^b$ should be vanishing up to a reasonable numerical
error~\cite{Donos:2013cka}.} and is consistent with the arguments
presented in ref.~\cite{Donos:2013cka}.


\section{Conclusion and discussions}
\label{sect:conclusion}
In this paper we studied a holographic p-wave superconductor model
in a four dimensional  Einstein-Maxwell-complex vector field theory
with a negative cosmological constant. The complex vector field
$\rho_\mu$ is charged under the Maxwell field. Taking the back
reaction of matter fields into consideration, we managed to
construct hairy black hole solutions which satisfy all asymptotic
conditions. We found the model presents a rich phase structure
controlled by the mass $m$ and charge $q$ of the vector field
$\rho_\mu$. We investigated possible phase transitions in detail. It
turns out that there exist zeroth order, first order and second
order phase transitions in this model. Hairy black holes were also
found in the unusual higher temperature range $T>T_n$, which always
have free energy higher than the normal phase. The phase diagrams in
terms of the temperature and charge were constructed.

Our numerical calculation suggests the existence of a critical $m^2$
denoted as $m_c^2=0$.  When $m^2>m_c^2$, we have a second order
phase transition from the normal phase to the condensed phase for
the weak back reaction case. This transition becomes a first order
one as we increase the strength of the back reaction. The transition
temperature $T_c$ decreases as we decrease the value of $q$, which
means that the increase of the back reaction makes the transition
more difficult. When $m^2<m_c^2$, the thermodynamic behavior of the
system changes a lot. Starting from the high temperature region, one
can find the following transitions: For $q>q_\alpha$, the system
undergoes a second order phase transition from the normal phase to
the condensed phase at $T_2$ and as the temperature decreases to
$T_0$, there is a zeroth order transition back to the normal phase;
For $q_\beta<q<q_\alpha$, the system first undergoes a first order
phase transition from the normal phase to the condensed phase at
$T_1$, then at the lower temperature $T_0$, it comes back to the
normal phase by a zeroth order transition; For $q<q_\beta$, we can
only get hairy black hole solutions that are always subdominant in
the free energy referred to as ``retrograde condensation". Here the
concrete values of $q_{\alpha}$ and $q_{\beta}$ depend on the mass
squared $m^2$ of the vector field $\rho_{\mu}$.

It was argued in ref.~\cite{Aprile:2010yb} that the holographic free
energy can be  thought of as a sort of generalized version of
Landau-Ginzburg free energy. In Landau-Ginzburg theory, it is
usually assumed that the quadratic term depends on the temperature
linearly while the fourth order term is not strongly temperature
dependent. In our present study, we found the behaviors deviating
from the mean field theory. In the context of Landau-Ginzburg
theory, such deviation would be a sign of an unusual temperature
dependence of the higher order terms. It should be stressed that our
model is dual to a strongly coupled system. A priori the dual system
does not obey the usual assumption for the free energy. It is in
principle possible that there is some new temperature scale at which
the coefficients of higher order terms change their signs, giving
rise to non-standard phase transitions in the framework of
Landau-Ginzburg theory.

Our study can be straightforwardly
generalized to  the higher dimensional case and other gravitational
backgrounds, such as the AdS soliton backgrounds which can mimic the
superconductor/insulator phase transition~\cite{Nishioka:2009zj}. In
a recent paper~\cite{Cai:2013kaa}, we studied the effect of an applied magnetic field
effect on the AdS soliton background, and found that
 the magnetic field can induce the AdS soliton instability due to the non-minimal
 coupling of the vector field and the background magnetic field.
 By comparing our complex vector field model to
the SU(2) p-wave model with a constant non-Abelian magnetic field, we found
that the SU(2) p-wave model can be recovered by the restriction $m^2=0$ and
$\gamma=1$ in our model with the ansatz in ref.~\cite{Cai:2013kaa}.
 It suggests that in some sense,
the charged vector model is a generalization of the SU(2) p-wave model to
the case with a general mass squared $m^2$ and gyromagnetic ratio
 $\gamma$ for the vector field.  Due to the adjustable parameter
$m$, we can see in this paper that our model shows a much richer
phase structure than the SU(2) p-wave model, thus can be used to describe
more phenomena in dual strongly coupled systems.

In the present paper, we limited ourselves to a simple case with
$\rho_x$ non-vanishing only.  In principle, in order to understand
the full phase structure of the model at fixed chemical potential,
one should search for the dominant thermodynamic configuration not
only in this given sector but in a more general setup, especially
turning on the temporal component of the charged vector $\rho_t$.
This would of course be much more involved, since one should search
for the hairy black hole configuration with the least free energy
among all possible configurations. We will leave this issue for
further study.

As a phenomenological approach, we consider the model as a p-wave
superconducting (superfluid) one.  Indeed, this toy model could be
applicable in a wide variety of condensed matter systems and beyond.
Indeed, as we have discussed in ref.~\cite{Cai:2013pda}, it may also
be revelent for holographically mimicking the phenomenon that the
QCD vacuum undergoes a phase transition to an exotic phase with
charged $\rho$-meson condensed in a sufficiently strong magnetic
field~\cite{Chernodub:2010qx,Chernodub:2011mc}.

As we have mentioned above, according to the symmetry of the
macroscopic wave function or  condensate of Cooper pairs in the real
superconducting materials, the superconductor can be classified by
s-wave, p-wave, d-wave and so on. The holographic s-wave model has
well studied (especially with back reaction) in the literature.
Adopting the present p-wave model, it is quite interesting to study
holographic models with multiple superconducting order parameters,
including the competition or coexistence between s-wave order and
p-wave order or between two p-wave orders.~\footnote{It might be
difficult to combine the s-wave model~\cite{Hartnoll:2008vx} and the
SU(2) p-wave model~\cite{Gubser:2008wv} in one theory, since the
order parameter in the SU(2) model is charged under a U(1) subgroup
of the Yang-Mills field and this U(1) subgroup can not play the role
of the gauge group for the s-wave model in a natural manner. The
current p-wave model is charged under a U(1) gauge group which also
can be naturally taken as the gauge group of the s-wave order. The
competition and coexistence between two s-wave orders were first
studied in ref.~\cite{Basu:2010fa} in the probe limit, and in
ref.~\cite{Cai:2013wma} with back reaction. More recently,
ref.~\cite{Nie:2013sda} constructed a model with a scalar triplet
charged under a SU(2) gauge field, there the s-wave order and p-wave
order can coexist.} We will leave all these issues for further
study.

Finally we like to mention that with suitable parameters $m$ and $q$, our model shows
the normal/superconducting/normal phase transition (see figure~\ref{phasediagn})
as one lowers the temperature continuously. Such a phase transition is called reentrant phase
transition  in the literature~\cite{report}. The reentrant phase
transition usually happens in the binary and multicomponent liquid mixtures. But it is interesting
to note that such a phase transition also appears in some superconducting materials, for example,
 granular ${\rm BaPb_{0.75}Bi_{0.25}O_3}$ compound~\cite{granular} and cuprate superconductors~\cite{cuprate}.
 Thus it would be of some interest to see whether our model is relevant to these superconducting phase transitions.

\section*{Acknowledgements}

This work was supported in part by the National Natural Science
Foundation of  China (No.10821504, No.11035008,
No.11205226,No.11305235, and No.11375247), and in part by the
Ministry of Science and Technology of China under Grant
No.2010CB833004.

\appendix

\end{document}